# Sleep Stage Scoring Using Joint Frequency-Temporal and Unsupervised Features

Mohamadreza Jafaryani
Department of Electrical Engineering, Amirkabir University of Technology
Jafaryani@aut.ac.ir

Saeed Khorram
Department of Electrical Engineering, Amirkabir University of Technology
Khorram.saeed@aut.ac.ir

Vahid Pourahmadi
Department of Electrical Engineering, Amirkabir University of Technology
v.pourahmadi@gmail.com

Minoo Shahbazi
School of Allied Medical Sciences, Tehran University of Medical Science
M_shahbazi@razi.tums.ac.ir

*Abstract*— Patients with sleep disorders can better manage their lifestyle if they know about their special situations. Detection of such sleep disorders is usually possible by analyzing a number of vital signals that has been collected from the patients. To simplify this task, a number of Automatic Sleep Stage Recognition (ASSR) methods have been proposed. Most of these methods use temporal-frequency features that have been extracted from the vital signals. However, due to the non-stationary nature of sleep signals, such schemes are not leading an acceptable accuracy. Recently, some ASSR methods have been proposed which uses deep neural networks for unsupervised feature extraction. In this paper, we proposed to combine the two ideas and use both temporal-frequency and unsupervised features at the same time. To augment the time resolution, each standard epoch is segmented into 5 sub-epochs. Additionally, to enhance the accuracy, we employ three classifiers with different properties and then use an ensemble method as the ultimate classifier. The simulation results show that the proposed method enhances the accuracy of conventional ASSR methods.

## I. INTRODUCTION

Sleep is considered as a paramount physical and mental state of humankind. Not only disorders in sleep will affect people's physical health, but also it may lead to some mental illnesses which can deeply impact one's quality of life. The observation of sleep stages is the primary tool that aids physicians to diagnose sleep disorders e.g., sleep apnea, narcolepsy. To that end, monitoring and recording activities during sleep period is of high importance in every aspect. In general, experts and sleep specialists determine the stages of sleep by visually analyzing the different epochs of the vital signals recorded from the patient. To simplify this process (as human-based monitoring might be very time consuming task), many efforts have been put into work reaching an automatic approach to address this problem. Moreover, sleep scoring is considered as a classification problem of vital signals received from the patient. Amongst the vital signals, the EEG signal is the most significant one and it has been used in this paper.

The standard method for sleep staging is based on the criteria proposed by Rechtschaffen & Kales [1] (R&K). According to this method, human's sleep is divided into 6 essential stages: Awake, Non Rapid Eye Movement(NREM) stage 1, NREM stage 2, NREM stage 3, NREM stage 4, Rapid Eye Movement(REM). However, in another method proposed by American Academy of Sleep Medicine (AASM), NREM stages 3 and 4 from the standard R&K were merged into the new stage referred to as Slow Wave Sleep (SWS) [2].

Different approaches have been proposed to extract features from the EEG signals. Amongst them, temporal [3-4] and frequency features [5-7] were drawn the most attention. The shortcoming of these schemes is that the EEG is regarded as a non-stationary signal. Henceforth, the conventional temporal and frequency features couldn't capture all the information and so does not lead to a great accuracy of classification. In recent years, there has been a resurgence of interest in deep learning systems. Deep Belief Networks (DBNs) were widely used to extract features from vital signals for sleep stage classification, and the results have been satisfactory [8].

In this paper, we proposed using temporal-frequency and unsupervised features of the EEG signal in a hybrid form. The joint feature set can boost systems' ability in sleep staging comparing with when they're restricted only to one of mentioned features. Moreover, in our research, three classifiers: Gaussian Process [9], Random Forest [10], and Hidden Markov Model (HMM) [11] have been used, and the results were evaluated.

Furthermore, in order to increase the time resolution, we segmented each standard epoch to 5 sub-epochs. The ultimate decision about the primary epoch is made based on the classification results that we get for each sub-epoch. Additionally, to enhance the classification results, we also proposed to utilize an ensemble method in which all three classifiers' results are considered in decision making, simultaneously.

This paper is organized as follows. First, in section II, DBN, as an unsupervised feature extractor, is introduced. In section III, blocks of the proposed framework for an ASSR system is reviewed. Section IV presents the experimental setups and simulation results, and finally, section V gives the conclusion.

## II. DEEP BELIEF NETWORKS

Deep Belief Network (DBN) is a multi-layer generative graphical model that can learn the probability distribution of the input data via its hidden layers [12]. Each DBN, as depicted in Fig. 1, is made up of stacked Restricted Boltzmann Machines (RBMs) in which a hidden layer of an RBM act as the visible layer of the subsequent layer. The connections between the hidden and visible layers are bidirectional, and units of each layer are fully connected to ones of the other. Note, that in RBMs, units of one layer are not allowed to have connections to each other. Each configuration of the RBM can be followed by a joint probability over hidden and visible variables [13]:

$$P(v,h;\theta) = \frac{e^{-E(v,h)}}{\sum_{u,g} e^{-E(u,g)}} \qquad (1)$$

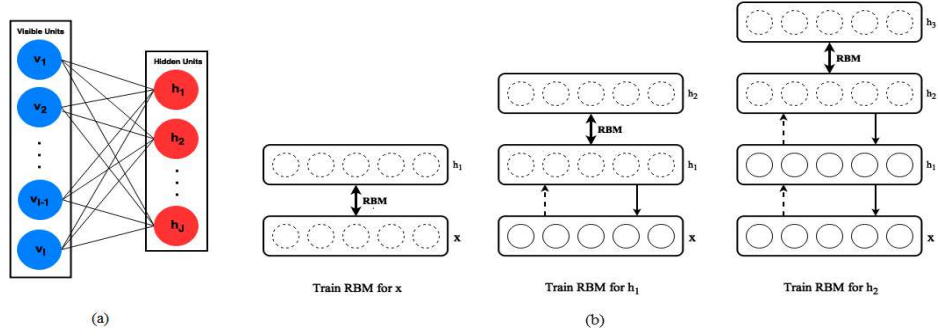

Fig.1. Graphical Representation of (a) RBM and (b) DBN

The energy function is defined as follows:

$$E(v,h) = -a^T v - b^T h - h^T W v \qquad (2)$$

Where a and b are the bias vectors of the visible and hidden layers, respectively, and W represents the weights connecting the units of the hidden and visible layers.

Due to the independency of the layers, the following conditional probabilities can be readily calculated as:

$$P(h_j = 1 | h; \theta) = \sigma\left(\sum_{i=1}^{I} w_{ij} v_i + a_j\right) \qquad (3)$$

$$P(v_j = 1 | h; \theta) = \sigma\left(\sum_{j=1}^{J} w_{ij} h_j + b_i\right) \qquad (4)$$

Where

$$\sigma(x) = \frac{1}{1 + e^{-x}} \qquad (5)$$

By taking gradient of log $P(v|\theta)$, the following rule for updating the RBM weights can be reached:

$$\Delta w_{ij} = E_{data}(v_i h_j) - E_{model}(v_i h_j) \qquad (6)$$

where $E_{data}(v_i h_j)$ is the expectation of the training set, and $E_{model}(v_i h_j)$ represents the expectation computed form the probability density which is introduced by the model. The computation of $E_{model}(v_i h_j)$ is intractable. Henceforth, some approximations are applied to tackle this problem. The Contrastive Divergence (CD) method, which is an approximation of the gradient, is extensively used to train RBMs. In this method, the $E_{model}(v_i h_j)$ is replaced by a sample form Gibbs sampler [14].

Training a DBN follows a layer-wise procedure that the first layer is trained, then the hidden values of the first layer become the input data of the next layer. The same follows until all layers are trained. Moreover, to fine-tune the weights of a generative DBN network, unsupervised backpropagation algorithm [15] can be applied.

DBNs with consecutive decreasing number of nodes in each layer can be used to learn unsupervised features. To put it differently, the following generated features are an abstraction of high dimension data. Such high level abstraction has been extensively utilized in applications like classification.

### III. PROPOSED AUTOMATIC SLEEP SCORING FRAMEWORK

As depicted in Fig. 2, the proposed framework for sleep stage classification is comprised of "preprocessor", "time-frequency feature extractor", "unsupervised feature extractor", and "classifier" subsystems. In the following, we will describe each block in more details.

*A. Preprocessing unit*

The collected raw data (EEG signal) is usually very noisy and it is essential to polish this data before we give it to the subsequent ASSR blocks. To clean the data, first, a notch filter at 50Hz is applied to the EEG signal aiming to reduce the disturbance regarding the power lines and electronic equipment affection. Additionally, as EEG signal's frequency is ranged from 0.3 to 32 Hz, with a corresponding band pass Filter, the redundant frequencies are eliminated. In the next step, the sampling frequency of the filtered EEG signal is converted (Up-sampled or Down-sampled) to 64 Hz. Moreover, to augment the time resolution, the standard epoch consisting of 30-second samples of EEG signals is segmented to 5 sub-epochs of 6 seconds each. Then, epochs having stage-change in their neighborhood is removed. Although the elimination of transition-epochs may treat ASSR system with less training data, the systems' classification ability is boosted due to the higher confidence in the training data.

Finally, the training data for each class is balanced. In other words, we only keep a number of training data such all classes (sleep stages) have an equal number of epochs comparing others, i.e. all classes will have the same number of epochs as the one with the least epochs.

*B. Time-Frequency Features extractor unit*

Having the polished data, this block extracts 14 features from both the time and frequency domain of each epoch.

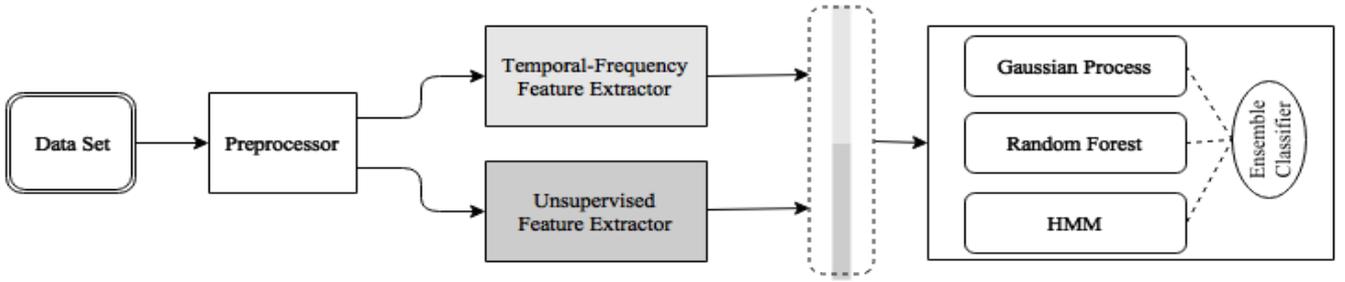

Fig.2. The Proposed Block Diagram of ASSR System

*1) Relative spectral power:*

5 features expressing the spectral activity of the EEG signal in the common frequencies bands are calculated through the ratio of power of the signal in a single frequency band to over total power of the signal as follows:

$$x_{p_{rel}}(f) = \frac{x_p(f)}{\sum_{f=f_1}^{f_5} x_p(f)} \quad (7)$$

Where $x_p(f)$ is the power of the signal x in the frequency band f. Moreover, the 5 mentioned frequency bands are δ [0.5–4 Hz], θ [4–8 Hz], α [8–13 Hz], β [13–20 Hz], and γ [20–64 Hz].

*2) Entropy:*

The entropy is computed from the histogram of the signal in each epoch using:

$$entr = \sum_{j=1}^{N} \frac{n_j}{n} \ln \frac{n_j}{n} \quad (8)$$

In the above equation, n represents the number of the total samples in one epoch. N and $n_j$ are the bin numbers used for calculating the histogram and the number of samples in $j$th bin, respectively.

*3) Harmonic parameters:*

Harmonic parameters of a signal include central frequency, bandwidth, and the spectral value at the central frequency which are calculated as follows [16]:

$$f_c = \sum_{f_l}^{f_h} f p_{xx}(f) \quad (9)$$

$$f_\delta = \sqrt{\frac{\sum_{f_l}^{f_h} (f-f_c)^2 f p_{xx}(f)}{\sum_{f_l}^{f_h} p_{xx}}} \quad (10)$$

$$s_{f_c} = p_{xx}(f_c) \quad (11)$$

Where $P_{xx}(f)$ stands for PSD of EEG signal.

*4) Hjorth parameters:*

Hjorth parameters provide the dynamic temporal features of EEG signal which are calculated using the variance of the signal x, first derivative (x') and second derivative (x") of the signal [17]:

$$\text{Activity} = \text{var}(x) \quad (12)$$

$$\text{Mobility} = \sqrt{\frac{\text{var}(x')}{\text{var}(x)}} \quad (13)$$

$$\text{Complexity} = \sqrt{\frac{\text{var}(x'')\text{var}(x)}{\text{var}(x')^2}} \quad (14)$$

*5) Skewness and kurtosis*

Skewness and kurtosis which respectively, describe measures of symmetry and the flatness of a distribution are defined as:

$$\text{skew} = \frac{m_3}{\sqrt{m_2}m_2} \quad (15)$$

$$\text{kurt} = \frac{m_4}{m_2 m_2} \quad (16)$$

$$m_k = \frac{1}{n}\sum_{i-1}^{n}(x(i)-\bar{x})^k \quad (17)$$

### C. Unsupervised Features extractor unit

As mentioned in section II, to extract unsupervised features from the raw data, we can benefit from DBNs by constructing a network in which the layer size is decreasing consequently as going forward in the network (Fig. 3). The output is a high level representation of the EEG signal.

Training a DBN (Determining the weights of the network) has two primary parts. First, the weights of the networks are pre-trained by using training data in an unsupervised manner. Second, the same weights are fine-tuned through the unsupervised back propagation.

Further, one of the paramount issues with respect to the training A DBN is the number of the layers as well as the number of neurons in each layer. As there is no specific way determining those, in this paper, we proposed to use the following method to attain a proper configuration. First, using 90% of the training data, various DBNs are trained with different configurations. Afterwards, the reconstruction error of the each DBN is evaluated on the rest of the training set. The network setting with the least reconstruction error is our selected DBN configuration. The selected DBN is then trained with the whole training set, and the ultimate weights are determined. Now, by having each epoch of the train or the test set, unsupervised features regarding each epoch are obtained as the output of the network. These features will be used in for sleep stage classification.

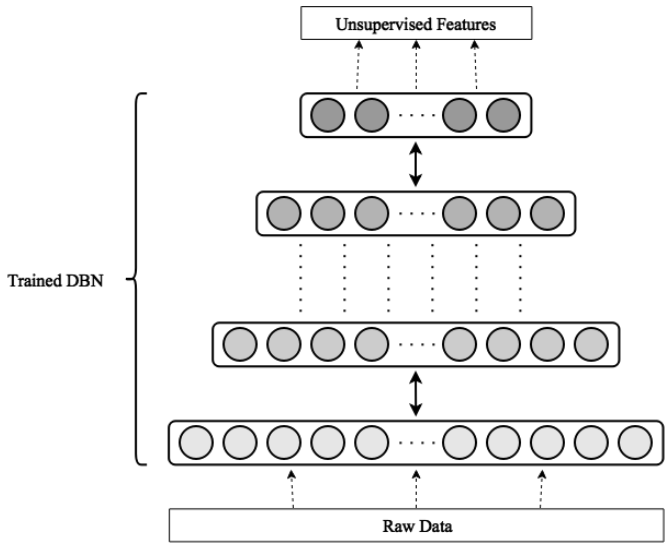

Fig.3. Unsupervised Feature Learning Block

*D. Classifier unit*

In the last block of the proposed ASSR system, both unsupervised and temporal-frequency feature vectors are combined in order to have a joint feature vector for each individual epoch. The combined feature vector is regarded as the input vector of the classifier. Moreover, in this paper, three classifiers, namely, the Gaussian process, the random forest, and the HMM have been used. Clearly, each of these classifiers have different advantages and drawbacks. Gaussian process is a non-parametric method that comes along with high accuracy; however, it roughly takes lots of time for training the classifier. Comparatively, random forest is considered as an ensemble-of-trees method which outperforms other classifiers regarding its implementation ease. Additionally, the random forest also benefits from its high classification speed. Furthermore, to count in the temporal information, HMM, as a dynamical model, has been used for classification.

## IV. EXPRIMENT AND RESULTS

The publicly available dataset used in this paper contains 25 acquisitions form suspected adult subjects with sleep-disorders. Subjects were randomly selected from the patients referred to the Sleep Disorder Clinic at St Vincent's University Hospital, Dublin. The subjects were 21 Male and 4 Females, with average age: 50 years, average height: 173 cm, and average BMI: 31.6; further details are available on PhysioNet [18]. Several vital signals are included in the dataset, but for this paper, only a single-channel EEG Signal (C3-A2) was selected for further analysis. It is clear that this work can be extended such that other vital signals or a combination of them are used as well. In this dataset, the stages were labeled as: Awake, REM, Stage 1, stage 2, stage 3, stage 4, Artifact, and Indeterminate which the last two were removed at the pre-processing stage. From the 25 acquisitions, 10 were randomly selected for training of the ASSR system, and 3 other patients were selected from the remaining ones to test and evaluate the systems' performance. By selecting only 10 acquisitions for the training phase, we were intended to show that the proposed ASSR system can perform competently with limited dataset.

As mentioned in section III.B, to extract unsupervised features from the raw data, different configurations have been tested, and the one with the least Mean Square Reconstruction Error (MSRE) was selected. To obtain a network with 15 or 20 unsupervised features, the 6 and 7 layer DBNs were followed by an acceptable MSREs. Table I shows 4 different configurations with corresponding MSREs. The best one (at the last row of the table) was selected to extract 15 unsupervised features. (The four mentioned configurations were just examples and many others have been tested.)

Table I. MSRE for various DBN configurations

| LAYER CONFIGURATION | MSRE |
|---|---|
| 300 → 225 → 150 → 50 → 40 → 30 → 20 | 0.034 |
| 200 → 150 → 80 → 50 → 35 → 25 → 20 | 0.0077 |
| 200 → 100 → 50 → 25 → 20 → 15 | 0.0049 |
| 150 → 75 → 35 → 25 → 20 → 15 | 0.0025 |

Furthermore, to obtain a joint feature vector for each epoch, 14 temporal-frequency features as mentioned in section III.C were combined with corresponding unsupervised features. The new joint feature vector was comprised 29 features for each epoch. Temporal-frequency and unsupervised features were examined both separately and jointly with three different classifiers. As illustrated in Fig. 4, proposed method based on combining the features outperforms other methods in which only one type of (Temporal-frequency and Unsupervised) features is utilized. Moreover, amongst the classifiers, as expected, the Gaussian process surpasses other classifiers.

To enhance the classification accuracy, the ensemble method is used. In this method, the majority vote scheme will run over the data, and when it fails to reach a unique answer, the results from the Gaussian process classifier are considered as the final answer.

As illustrated in Table II, the accuracy of 89.47% was reached for using the proposed method in the paper. It is also perceived that the easiest stages to recognized are the SWS and REM, respectively. The hardest one with the least accuracy is NREM stage 1.

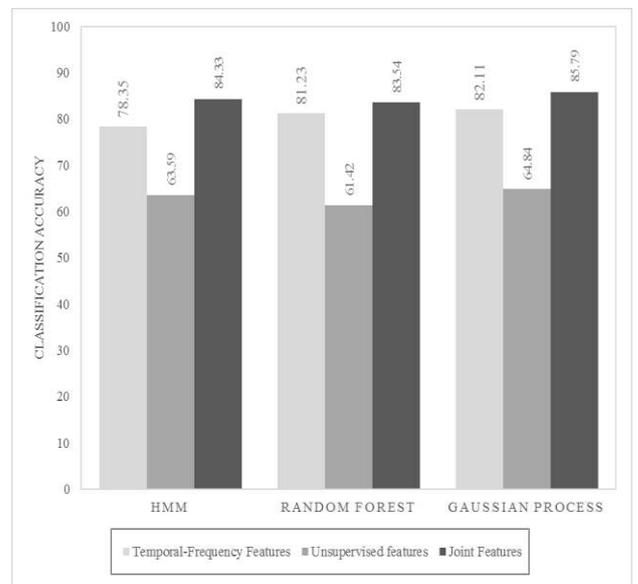

Fig.4. The classification accuracy using three different classifiers for temporal-frequency, unsupervised, and proposed joint features separately.

Table II. The confusion matrix for ensemble method

| % Classified | Awake | NREM1 | NREM2 | SWS | REM |
|---|---|---|---|---|---|
| Awake | **88.23** | 6.32 | 1.5 | 0.18 | 3.77 |
| NREM1 | 10.67 | **68.31** | 11.44 | 0.90 | 8.68 |
| NREM2 | 1.12 | 6.81 | **86.21** | 4.49 | 1.37 |
| SWS | 0.24 | 0.34 | 5.60 | **93.3** | 0.52 |
| REM | 1.83 | 2.48 | 4.51 | 0.93 | **90.25** |

**Total accuracy: 89.47**

## V. CONCLUSION

In this paper, sleep stages have been classified using both unsupervised and temporal-frequency features. The proposed method takes advantage of the strong ability of DBNs in modeling non-linearity in the data which can solve the major issue of conventional EEG-based classifiers. Also, taking into account the temporal-frequency features as well as using an ensemble method leads to a higher confidence in classification. Our results showed that the proposed method is able to perform the sleep staging task with an average accuracy of 89.47%, only using a single EEG channel (C3-A2) and limited training data.

## VI. REFRENCES